\documentclass[aps,pra,twocolumn,showpacs]{revtex4-1}

\usepackage{amssymb,amsmath,amstext}                
\usepackage{graphicx}                                               
\usepackage{epstopdf}                                               
\usepackage{color}                                                     
\usepackage{bm}                                                        
\usepackage{appendix}                                              
\usepackage[utf8]{inputenc}
\usepackage{bbold}
\usepackage{bbm}
\usepackage{ulem}
\usepackage[dvipsnames]{xcolor}

\usepackage{latexsym}
\usepackage[colorlinks=true,citecolor=blue,linkcolor=magenta]{hyperref}

\usepackage{filemod}

\def\be{\begin{equation}}
\def\ee{\end{equation}}
\def\bg{\begin{equation}\begin{gathered}}
\def\eg{\end{gathered}\end{equation}}

\begin{document}

\title{Spin-charge separation and many-body localization}

\author{Jakub Zakrzewski}
\affiliation{Instytut Fizyki imienia Mariana Smoluchowskiego, Uniwersytet Jagiello\'nski,   ulica Profesora Stanis\l{}awa \L{}ojasiewicza 11, 30-348 Krak\'ow, Poland \\ and
Mark Kac Complex
Systems Research Center, Uniwersytet Jagiello\'nski, Krak\'ow,
Poland. }
\email{jakub.zakrzewski@uj.edu.pl}

\author{Dominique Delande}
\affiliation{Laboratoire Kastler Brossel, UPMC-Sorbonne Universit\'es, CNRS, ENS-PSL Research University, Coll\`ege de 
	France, 4 Place Jussieu, 75005 Paris, France}
\affiliation{
	Instytut Fizyki imienia Mariana Smoluchowskiego, 
	Uniwersytet Jagiello\'nski, ulica Profesora Stanis\l{}awa \L{}ojasiewicza 11, 30-348 Krak\'ow, Poland}


\begin{abstract}
We study many-body localization for a disordered chain of spin 1/2 fermions. In [Phys. Rev. B \textbf{94}, 241104 (2016)], when both down and up components
are exposed to the same strong disorder, the authors observe a power law growth of the entanglement entropy that suggests that many-body localization is not complete; the density (charge) degree of freedom is localized, while the spin degree of freedom is apparently delocalized. We show that this power-like behavior is only a transient effect and that, for longer times, the growth of the entanglement entropy is logarithmic in time. At the same time, the decay of the spin correlation slows down, showing that the spin transport is strongly reduced at long time. The dynamics 
of the system is quite similar to the one of many-body localized systems. Because of the limited duration of the numerical experiment, it is not possible to determine whether the system is truly many-body localized or diffusing very slowly at long times.

\end{abstract}
\date{\today}

\maketitle
\section{Introduction}

Despite several years of intensive investigations, many-body localization (MBL) remains a very active field of research with hundreds of papers appearing per year on the subject. Early seminal works \cite{Basko06,Oganesyan07,Znidaric08} are already supplemented  by several reviews (see e.g. \cite{Huse14,Rahul15} as well as papers in a topical issue of Annalen der Physik~\cite{AnnPhys:MBL:2017}).  The popularity of the subject seems to be due to the fact that MBL is a counterexample of a basic hypothesis of statistical physics: complex many-body interacting system should thermalize. For large isolated systems, the  eigenvector thermalization hypothesis (ETH) \cite{Deutsch91,Srednicki94} indicates that any {\it local} observable should thermalize losing, during the temporal evolution, most of the detailed information on the initial state. One of the key features of MBL is the opposite behavior: the system remembers its initial state as some local observables do not tend at long times towards 
their thermal average values.

The ``standard theoretical model'' of MBL is the spin-1/2
 Heisenberg chain of length $L$ with Hamiltonian (here written for open boundary conditions):
 \be
\label{Hsta}
\hat{\mathcal{H}}_\mathrm{spin}= J \sum_{i=1}^{L-1}\vec{S}_i\cdot\vec{S}_{i+1}
+ \sum_{i=1}^L h_i S_i^z
\ee
which, for random uniform $h_i\in[-H,H]$, shows a transition from an ergodic to MBL behavior for a sufficiently strong disorder \cite{Huse14}.
Using a Jordan-Wigner transformation, one can map the spin model to a system of interacting spinless fermions in a lattice~\cite{Lieb61}. Since, at a given site, a single fermion occupation is allowed, the interaction is possible only between spins in nearest sites. The experiment, on the other hand, favors spinful fermions with on-site interactions \cite{Schreiber15,Kondov15,Bordia16}. The Hamiltonian routinely realized in experiments reads:
\be
\label{H_FH}
\hat{\mathcal{H}} 
= -t_0 \sum_{\langle i,j \rangle,\alpha} (c_{i,\alpha}^\dagger c_{j,\alpha}+ h.c.)+  U\sum_{i}n_{i,\uparrow}n_{i,\downarrow}
+ \sum_{i,\alpha}\mu_{i,\alpha}n_{i,\alpha},
\ee
where $t_0$ denotes the tunneling between sites (the same for up and down pointing fermions), $c_{i,\alpha}, c_{i,\alpha}^{\dagger}$  the annihilation and creation operators for $\alpha=\uparrow,\downarrow$ fermions, and $n_{i,\alpha}$ are the corresponding occupation number operators at site $i$.
$\mu_{i,\alpha}$ are (random or quasi-random) chemical potential values at different sites while $U$ is the strength of on-site interactions between
the two spin components. Instead of the occupation number operators, one may conveniently define the site density $d_i\equiv n_{i,\uparrow}+n_{i,\downarrow} $ as well as the site magnetization  $m_i\equiv n_{i,\uparrow}-n_{i,\downarrow} $. In experiments the chemical potential is due to the optical potential created either by a speckle \cite{Kondov15} or by an additional laser beam with frequency incommensurate with the frequency 
of the optical lattice~\cite{Schreiber15,Bordia16}. In both cases,
it is spin ($\alpha$) independent; the disorders for up-fermions and down-fermions are then perfectly correlated, or, in other words, it acts in the charge/density sector, not in the spin/magnetization sector.

While experiments confirmed a MBL-like behavior in this system, they concentrate on the so-called charge degree of freedom, i.e. the site-dependent density, $d_i$ and its time-dependent correlation function $D(t)\propto \sum_i\langle (d_i(t)-\bar d)(d_i(0)-\bar d)\rangle$ where $\bar d$ is the mean density, normalized such that $D(0)\!=\!1.$ In the MBL phase, the correlator fluctuates around a non-zero mean value while, in the ergodic phase, it decays to zero - the system loses the memory of its initial density profile.

This system has been recently studied in~\cite{Pr16}. Numerical simulations show that, when uncorrelated (independent) disorders are used for the
two spin components, the usual MBL scenario is observed, where both the charge and spin degrees of freedom are localized for a sufficiently strong disorder, on a rather short time scale, of the order of few characteristic time $t_0^{-1}.$ However, a  surprizing finding has also been presented in \cite{Pr16}. For {\it correlated} disorders, i.e. when spin-up and spin-down fermions feel the same disorder, on a short time scale (100 $t_0^{-1}$), the charge degree of freedom is apparently localized, but the spin is delocalized, as shown by the temporal decay of the correlations in the spin sector. Thus MBL in such a case is apparently not complete. Another hallmark of MBL~\cite{Znidaric08,Serbyn13a}, often used in numerical experiments, is the growth in time of the entropy of entanglement between the right and left half-chains, when the system is cut e.g. in its middle. 
In the MBL regime, the entanglement propagates very slowly
	in the system. It is numerically observed~\cite{Bardarson12,Luitz16} and understood from a renormalization group approach~\cite{Serbyn13a,Nanduri14,Potter15,Vosk15} that the entanglement entropy grows only logarithmically with time. In contrast, in a delocalized system where the transport is diffusive or subdiffusive, the entanglement entropy increases as a power law of time~\cite{Luitz16}. For the correlated disorder, Ref.~\cite{Pr16} presented the numerical evidence that the entanglement entropy growth obeys a power law (although with a small power), suggesting some delocalization. 

The aim of this paper is to show that a detailed numerical study of the very same system (with correlated disorders) at longer times leads to a different conclusion: the decay of spin correlations and the power law increase of the entanglement entropy observed in~\cite{Pr16} are transient effects at relatively short times; At long times, the decay of spin correlations slows down and the entanglement entropy grows logarithmically with time, while the charge remains localized. Altogether, the behavior is similar to the one of a fully MBL system. Note however that numerical data over a finite time cannot determine the ultimate fate of the system. It might be that both charge and spin are localized, or that the spin transport persists at long time - for example subdiffusively as suggested in ~\cite{Kozarzewski:subdiffusive:18} - or that even the charge eventually delocalizes.

\section{Results of the numerical simulations}

We use the same system \eqref{H_FH} already studied in Ref.~\cite{Pr16} with the same parameters $t_0\!=\!U\!=\!1$ and unit filling $\bar d=1$. We take $t_0^{-1}$, the inverse of the hopping amplitude, as the unit of time. The disorder is taken from
a uniform random distribution in the $[-W,W]$ interval with $W=16$ and the system size is $L=64,$ with open boundary conditions. For the temporal propagation, we use the Time-Dependent Block Decimation TEBD (t-DMRG) algorithm \cite{Vidal03,Vidal04,Schollwoeck11}, in a home-made implementation taking advantage of the conserved quantities, namely the total numbers of
up and down fermions. This makes it possible, for strong disorder, to go to significantly longer times than in~\cite{Pr16}. The initial state is chosen
as a product state with unit filling (half spins up and half spins down) with
randomly chosen occupied sites. We use the ''natural'' observables
used in~\cite{Pr16}: the charge/density correlator $D(t)$ defined above, 
and the spin/magnetization correlator, defined by $M(t) = B \sum_i\langle m_i(t)m_i(0)\rangle,$ with $B$ the normalization constant ensuring $M(0)=1$.
We also concentrate on the entanglement entropy $S=-{\rm Tr} \rho_A \ln \rho_A$ after splitting the chain (at its center) dividing the system into two half-chains $A$ and $B$. In the TEBD algorithm, the state of the 
system is at all times a matrix product state~\cite{Schollwoeck11}, 
and $S$ is simply computed by  $S = -\sum \lambda_i \ln \lambda_i$ where the $\lambda_i$ are the weights of the links between the two half-chains, see~\cite{Schollwoeck11} for details. The numerical data presented below are averaged over many realizations of the disorder and many randomly chosen initial states. 

\begin{figure}
\includegraphics[width=1.0\linewidth]{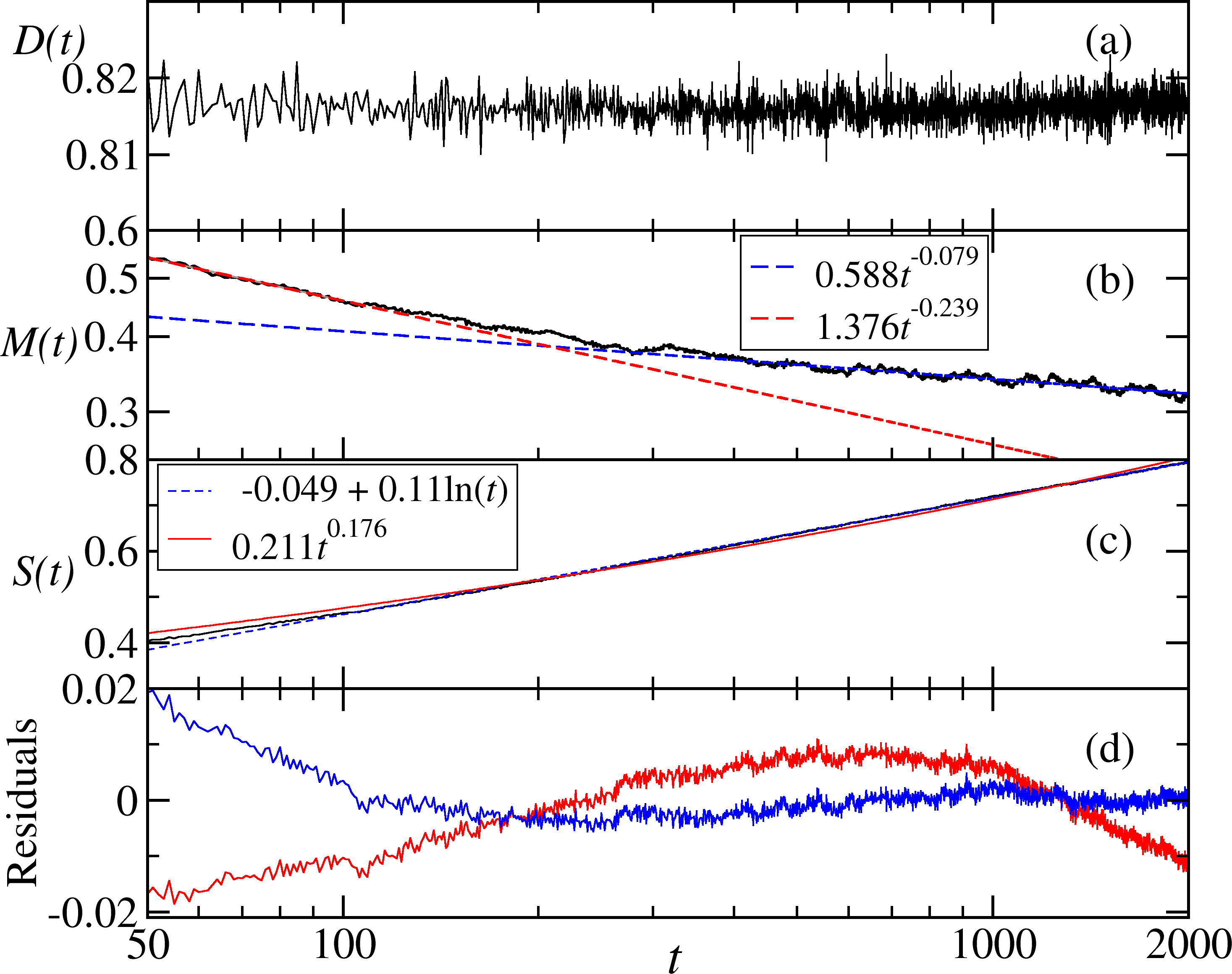}
\caption{Numerical results obtained for the disordered Hubbard chain, Eq.~(\ref{H_FH}), on a long time scale, up to $t\!=\!2000.$ Results are obtained by t-DMRG for a chain of length $L\!=\!64, U\!=\!t_0\!=\!1,$ disorder strength $W\!=\!16$ averaged over 237 realizations of the disorder. Panel (a) shows the charge correlator $D(t)$ which saturates to a finite value $\approx 0.815,$ indicating localization on this time scale. Panel (b) shows the decay of spin correlation $M(t)$ which is algebraic at short time, and clearly changes behavior around $t\!=\!300$ towards a slower decay. Panel (c): the entropy of entanglement (black solid line) grows logarithmically in time as shown by the corresponding fit (blue dashed line) which is clearly superior to the power law behavior (red solid line). That is further visualized in panel (d) showing the differences between the numerics and the fits. Both fits were performed in the full [50,2000] time interval. }
\label{fig:64_longtime}
\end{figure}

For relatively short time (up to 100), we confirm the results of~\cite{Pr16}, that is saturation of the charge correlation to a finite value, indicating localization, while the
spin correlator decreases algebraically $M(t)\propto t^{-\alpha},$ with $\alpha\approx 0.239.$ The entanglement entropy grows algebraically $S(t)\propto t^\sigma$ with $\sigma=0.20$, in good agreement 
with  \cite{Pr16} which observes $\sigma=0.18$~\footnote{We use the standard natural log in the definition of the entropy, while~\cite{Pr16} uses a base 2 logarithm; this results in an unimportant global multiplicative factor, and trivial differences in vertical scales, between \cite{Pr16} and the present work. }

On a longer time scale (up to $t\!=\!2000$), a different picture emerges. The results, presented in Fig.~\ref{fig:64_longtime}, show that the charge degree of freedom remains localized (no decay of the $D(t)$ correlator is observed on this time scale). The behavior of the spin correlator $M(t)$ very significantly changes around time $t\!=\!300$ where the decay slows down. The present data are insufficient to assess if it actually tends to a finite value at infinite time  -- which would indicate localization -- or slowly decays towards 0, which should be the case for diffusive or subdiffusive spin transport. The very slow decay at long times is too slow to determine whether it is algebraic, logarithmic or something else. A power law fit gives an exponent $\alpha=0.079,$ three times smaller in magnitude than in the short time range.  

The entanglement entropy gives a complementary information, see Fig.~\ref{fig:64_longtime}(c) and (d). While, in agreement with \cite{Pr16}, "the growth is better described by a power law" at relatively short time, the opposite is true at long times. The growth 
is much better fitted in the [50,2000] time interval by a logarithmic increase than by a power law.
This is further exemplified in panel (d) showing residuals (differences between numerical and fitted values). Both fits show deviations at short time,
but, for $t>300,$ the logarithmic growth is unambiguously and vastly superior.

Altogether, the growth of the entanglement entropy and the very slow decay of the spin correlation at long time are hints of localization in the spin sector. At longer times, not easily accessible with present computer resources, it could be that the spin correlator saturates at a finite value, restoring MBL for charge and spin (on different time scales). It could also be that the spin correlator continues to decay towards 0, indicating delocalization in this sector. This should be accompanied by an algebraic growth of the entanglement entropy, which is, however, not visible in our data. Whatever the ultimate fate is, it remains that something unclear takes place around time $t\!=\!300$, where spin transport slows down and entanglement entropy  starts to increase logarithmically. {This characteristic time scale is definitely much longer than the spin
localization time - of the order of few units - when independent
disorders are used for the two spin components~\cite{Pr16}.} Additional theoretical work is needed to understand this behavior.

\begin{figure}
	\includegraphics[width=1.0\linewidth]{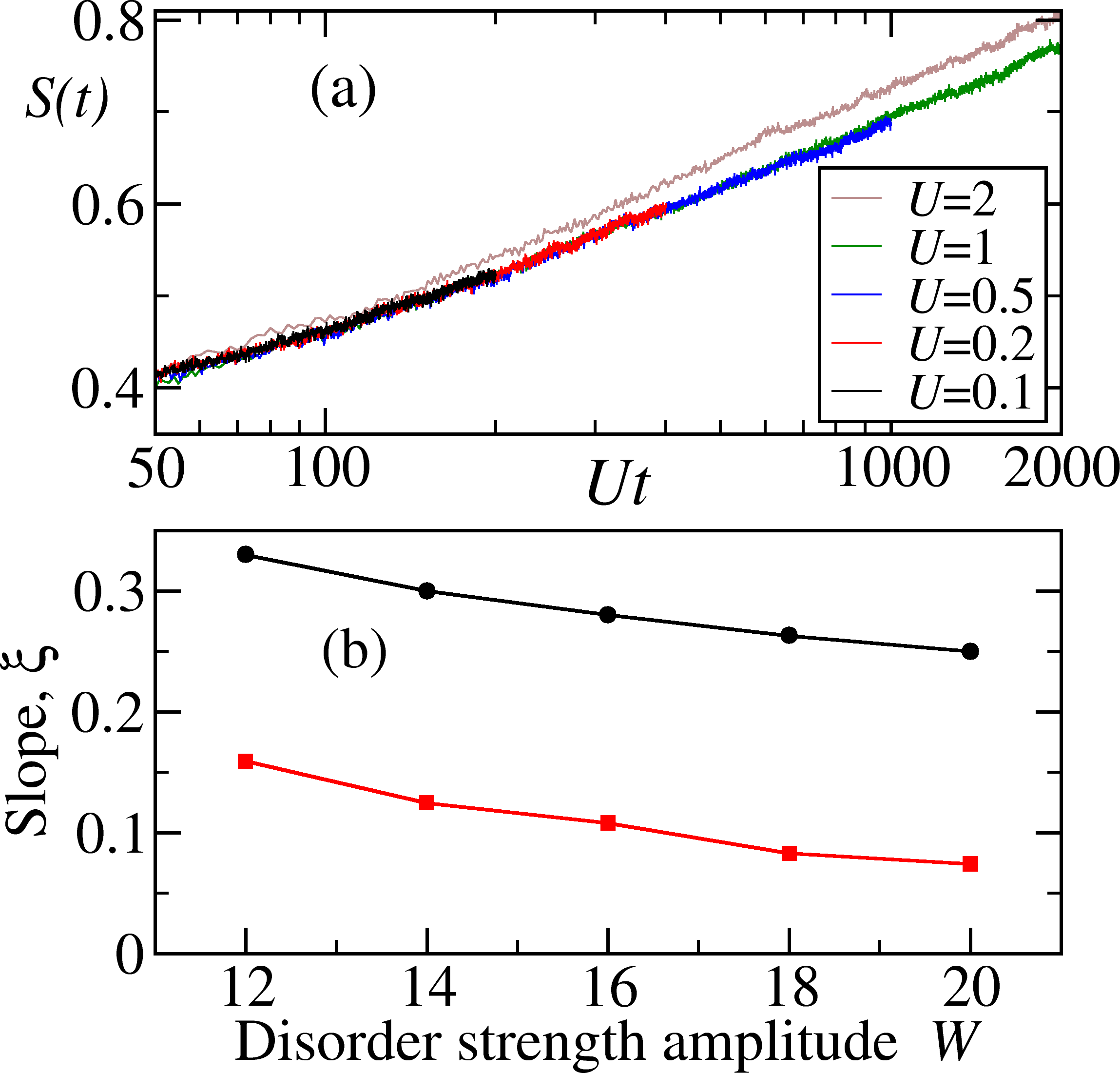}
	\caption{
		Dependence of the entanglement entropy with the interaction strength. The theoretical prediction in the MBL regime, adapted from~\cite{Serbyn13a} is that $S(t)$ is proportional to $\xi(W) \log(Ut)$, where $U$ is the interaction strength between spin up and spin down and $\xi(W)$ the non-interacting (Anderson) localization length. Panel (a): Results for the same $W=16$ and various $U$ values almost collapse on the same linear curve in linear-log scales, validating the $ \log(Ut)$ dependence. Panel (b): Comparison of the slope of $S(t)$ vs. $\log(Ut)$ for different values of $W$ with the noninteracting (Anderson) localization length $\xi(W)$.  }
	\label{fig:64_scaling}
\end{figure}

The logarithmic growth of the entanglement entropy in the many-body localized regime has been semi-quantitatively explained in Ref.~\cite{Serbyn13a}. Using the random-field XXZ model, a variant of Hamiltonian~(\ref{Hsta}), it predicts such a log dependence. When the interaction $V$ is small, the prediction is:
\begin{equation}
S(t) \propto \xi \log(Vt)
\label{eq:SPA}
\end{equation} 
where $\xi$ is the non-interacting localization length, i.e. the localization length in the Anderson localization regime. This model is simpler that the one we are studying, because there is only a single charge/density channel for transport. Our model has an additional spin channel where the temporal dynamics is very different. Is the simple analysis of ~\cite{Serbyn13a} relevant there? In order to answer this question, we have studied how the growth of the entanglement entropy at long time depends on two parameters: the interaction strength $U$ (which plays here the role of $V$ in ~\cite{Serbyn13a}), and the disorder strength $W$. Figure~\ref{fig:64_scaling}(a) displays $S(t)$ vs. $\log(Ut)$ for a fixed disorder strength $W\!=\!16$ and several values of $U.$ Remarkably, the curves almost collapse on a single curve, as predicted by Eq.~(\ref{eq:SPA}). A small offset is observed for the largest $U\!=\!2$ value, which is not unexpected, the prediction being valid in the small interaction limit. The slope is 
almost identical on all curves.
In Fig.~\ref{fig:64_scaling}(b), we plot the slope of $S(t)$ vs. $\log(Ut)$ as a function of $W.$ As expected, it decreases with increasing disorder $W.$ It follows more or less the decrease of the non-interacting (Anderson) localization length $\xi(W)$ computed for the same disorder. Note that the theoretical prediction (\ref{eq:SPA}) does not specify the proportionally factor, so that a factor 2 as observed in Fig.~\ref{fig:64_scaling}(b) is not inconsistent with \eqref{eq:SPA}. Thus, although the temporal dynamics of our system is rather complicated 
at long time, with an apparently localized charge channel and slow transport in the spin channel, the single channel prediction of~\cite{Serbyn13a} works surprizingly well.
It remains to understand why.

 \begin{figure}
 	\includegraphics[width=1.0\linewidth]{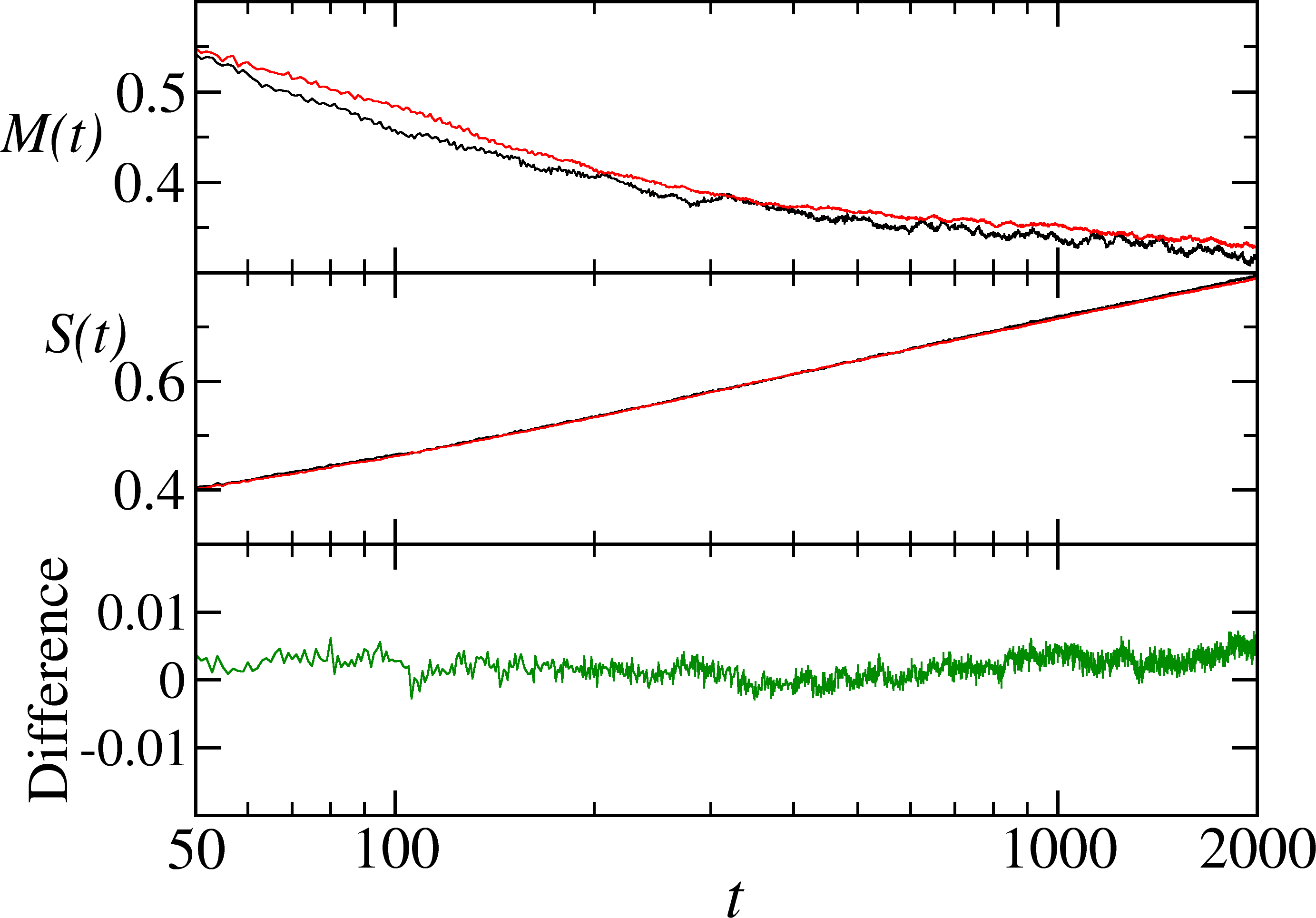}
 	\caption{
 	Convergence of the results obtained for two different bond dimensions $\chi$ in the t-DMRG algorithm~\cite{Schollwoeck11}. Black curves: $\chi=350$  with 237 disorder/(initial states) realizations (data of Fig.~\ref{fig:64_longtime}). Red curves:  $\chi=200$ with 979 disorder/(initial states) realizations. The lower panel shows the difference of entanglement entropies between the two cases studied.}
 	\label{fig:64_convergence}
 \end{figure}
 
We carefully checked that our numerical results are not impaired by finite computer resources. Fig~\ref{fig:64_convergence} compares the results obtained for two different values of the bond dimension in the t-DMRG algorithm~\cite{Schollwoeck11}: $\chi=350$ (value used
in Fig.~\ref{fig:64_longtime}) and $\chi=200$. The differences are small, signaling convergence of the results. Similarly, the time step and truncation errors are chosen sufficiently small.

\begin{figure}
	\includegraphics[width=1.0\linewidth]{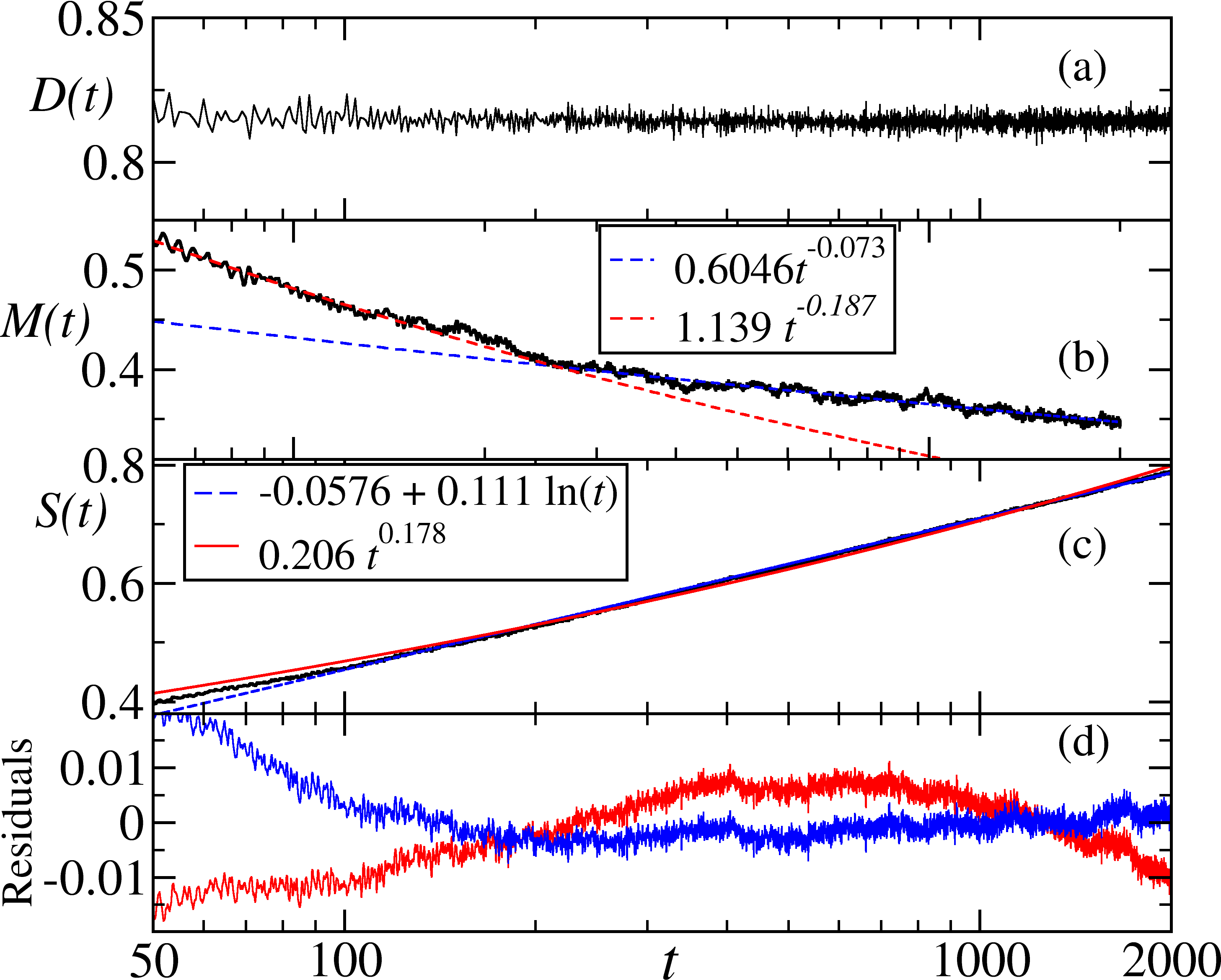}
	\caption{Numerical results for a disordered Hubbard chain of length $L=1400.$ The results are very similar to the ones shown for length $L=64$ in Fig.~\ref{fig:64_longtime}, proving that the observed behavior is not impaired by finite size effects. Panel (a) shows the saturation of the charge correlator, indicating localization in the charge sector. Panel (b) shows the decay of spin correlation $M(t)$ which again changes behavior around $t\!=\!300$ revealing a much slower decay at long times. Panel (c): Entanglement entropy $S(t).$ On a long time scale,
		the growth of the entanglement entropy is logarithmic in time, as clearly shown by the residuals of the fits shown in panel (d).}
	\label{fig:1400}
\end{figure}

We checked that our results are not dependent on the system size. We have studied the dynamics of a long chain of 1400 sites with 1400 fermions (initial state with 700 fermions up, 700 fermions down, on randomly distributed sites). Because of increased consumption of computer resources, we have only 10 realizations of disorder and the evolution was followed up to $t\!=\!2000$. As we have much more possibilities of cutting the system into two parts, the statistics  of entanglement entropy is still sufficient to draw unambiguous conclusion. The results  are shown in Fig.~\ref{fig:1400}. We again observe that for short time, the entanglement entropy growth may be fitted by a power law; This is not true at long time, where it grows logarithmically with time. The spin correlator, presented in panel (b), is very similar to the one observed for $L\!=\!64$ in Figure~\ref{fig:64_longtime}: it decays algebraically 
up to $t\approx 300$ with a slope $\alpha\approx 0.187.$ At longer time, the decay is much slower, so slow that it can be fitted equally well by several functional dependences. 
An algebraic fit in the [400,2000] range gives $\alpha \approx 0.073,$ very similar to the $L\!=\!64$ result.

\section{Conclusions}
To summarize, we have studied the long time dynamics of a chain of interacting spin 1/2 fermions exposed to a spatial disorder. For the
specific case where the disorder acts identically on the up and down components, it is observed that the charge/density degree of freedom is localized at strong disorder, in accordance with the usual MBL scenario.
In contrast, the spin degree of freedom is apparently delocalized at short time, as noticed in~\cite{Pr16}. However, at long time, the behavior changes drastically: The decay of the spin correlator slows down very significantly, and the entanglement entropy grows logarithmically with time. Whether the spin sector is truly localized or slowly diffusive
at very long time cannot be determined from the finite time dynamics.
Most probably, the separation of time scales between the charge and spin sectors is directly related to the fact that the disorder acts in the charge sector. The eventual localization in the spin sector is due to the slow transfer of information from the charge to the spin sector. 
One cannot exclude that, on an even longer time scale, MBL is destroyed, but this (exponential?) time scale is beyond the capabilities of our numerical simulations. 
Finally, let us note that a similar system was studied in~\cite{Mondaini15} using exact diagonalization for small system sizes $L\!=\!10$ and 12. The Hamiltonian was slightly different with additional next-to-nearest neighbor interaction, preventing a direct comparison. Moreover the boundary conditions broke the SU(2) symmetry of the model. In effect, for sufficiently strong disorder, both the charge and spin degrees of freedom were localized.

\bigskip

\acknowledgments
We thank Jan Major for providing us with noninteracting localization length data and Piotr Sierant for discussions and {suggestions about fits}. This work was performed with the support of EU via Horizon2020 FET project QUIC (nr. 641122). Numerical results were obtained with the help of PL-Grid Infrastructure. We acknowledge support of the National Science Centre, Poland via project No.2015/19/B/ST2/01028.

%

\end{document}